draft 3

# Semifluxon degeneracy choreography in Aharonov-Bohm billiards

M V Berry and S. Popescu

H H Wills Physics Labroatory, Tyndall Avenue, Bristol BS8 1TK, UK

**Abstract**

Every energy level of a charged quantum particle confined in a region threaded by a magnetic flux line with quantum flux one-half must be degenerate for some position of the semifluxon within the boundary B. This is illustrated by computations for which B is a circle and a conformal transformation of a circle without symmetry. As the shape of B is varied, two degeneracies between the same pair of levels can collide and annihilate. Degeneracy of three levels requires three shape parameters, or the positions of three semifluxons; degeneracy of $N$ levels can be generated by int$\{N(N+1)/4\}$ semifluxons. The force on the semifluxon is derived.

**Keywords**: Quantum, geometric phases, magnetic flux, eigenstates

**PACS numbers**: 03.65.Vf

# 1. Introduction

According to the von Neumann-Wigner non-crossing theorem [1], energy levels of quantum systems without symmetry typically do not intersect when a single parameter is varied. If the only symmetry is time reversal (T – here interpreted [2] as the Hamiltonian H being a real symmetric operator), degeneracies have codimension two: they typically occur as points in two-parameter families of spectra. The degeneracies are 'diabolical points' [3] where the intersections of the energy-level surfaces take the form of a double cone [4]. The same phenomenon occurs with the mathematically equivalent 'false time-reversal symmetry'[2], that is, invariance of the hamiltonian under any anti-unitary operator O satisfying $O^2=1$, because then there exists a basis for which H is real [2]. If there is no symmetry at all, degeneracy typically requires the variation of three parameters.

Our purpose here is to explore degeneracies in a class of systems with false time-reversal symmetry, namely Aharonov-Bohm (AB) billiards with half-odd integer quantum flux, that is semifluxons. AB billiards [5, 6] are planar domains, bounded by hard walls and pierced by a line of magnetic flux, within which a charged quantum particle is confined. The quantum flux, on which AB effects, in particular the spectrum of discrete energy levels, depend, is $\alpha$=charge×flux/$h$. AB billiards were introduced [5] as models for pure T-breaking in quantum chaology [7] – pure, because the pattern of classical orbits is unaffected by the flux. But when $2\alpha$ is integer these systems possess false T symmetry (section 2).

We will concentrate on the semifluxon case $\alpha=1/2$, which has many surprising properties [8]. For the two parameters $X$, $Y$ required to create degeneracies, we depart from previous studies [3, 9], in which $X$ and $Y$ parameterised the shape of the billiard boundary B. Instead, $X$ and $Y$ will now be coordinates of the position $\boldsymbol{R}=\{X,Y\}$ of the flux line within the fixed B.

By a gauge transformation, the wavefunction can be chosen to be real (but either double-valued or discontinuous) (section 2), and must possess a nodal line connecting the flux line with B. It is shown in Appendix A that each eigenstate exerts a force on the semifluxon, directed along the nodal line and tending to shorten it.

As a consequence of interaction between the AB phase and the geometric phase, degeneracies are not only possible but inevitable [8] (section 3): every energy level must degenerate an odd number of times for certain flux positions $(X, Y)$ inside the boundary. When B is a circle (section 4), the levels depend only on the distance $R = \sqrt{X^2 + Y^2}$ of the flux from the centre, and there are degeneracies at $R=0$. For non-circular boundaries, the degeneracies form interesting patterns, which we illustrate in section 5.

Degeneracies can themselves degenerate as further parameters vary, as discussed in section 6. Coincidence of three levels requires five parameters, for example three billiard shape parameters in addition to $X$ and $Y$, or the positions of three semifluxons rather than one (Y Aharonov, personal communication). Coincidence of two degeneracies between the same pair of levels requires only one additional parameter, and this is illustrated with a numerical example.

**2. Real and complex semifluxon wavefunctions**

The wavefunction $\psi(r;R)$ at position $r=\{x,y\}$ within B, for flux position $R=\{X,Y\}$ (figure 1) and energy $E=k^2$, satisfies

$$\left[ (\nabla_r - i\alpha A)^2 + k^2 \right] \psi = 0, \quad \psi = 0 \text{ for } r \text{ on B}, \tag{2.1}$$

where the imaginary unit i indicates T-breaking and the vector potential $A(r;R)$ satisfies

$$\nabla \times A = 2\pi\delta(r - R). \tag{2.2}$$

To get a gauge-invariant formulation, we write

$$\psi(r;R) = f(r;R) \exp\left\{ i\alpha \int_{r_0}^{r} A(r';R) \cdot dr' \right\}. \tag{2.3}$$

The transformed wavefunction satisfies the real equation and real boundary condition

$$\left[\nabla_r^2 + k^2\right]f = 0, \quad f = 0 \text{ for } \boldsymbol{r} \text{ on B}. \tag{2.4}$$

In this representation, the AB effect of the flux is transferred to a complex continuation rule, under which $f$ is no longer singlevalued: if $\boldsymbol{r}_C$ denotes the position after a closed path C starting at $\boldsymbol{r}$ (figure 1), the rule is

$$f(\boldsymbol{r}_C;\boldsymbol{R}) = f(\boldsymbol{r};R) \times \begin{pmatrix} \exp(-2\pi i\alpha) \text{ if C encloses the flux} \\ 1 \text{ if C does not enclose the flux} \end{pmatrix}. \tag{2.5}$$

(Alternatively, we can insist on singlevaluedness, in which case $f$ must be discontinuous across an arbitrary curve connecting the flux and B.)

For the semifluxon case $\alpha=1/2$ that we are interested in, the rule is

$$f(\boldsymbol{r}_C;\boldsymbol{R}) = -f(\boldsymbol{r};R) \text{ if C encloses the semifluxon}. \tag{2.6}$$

Now the continuation rule is real as well as the equation and boundary condition, so we can choose $f$ real. An immediate consequence is that $f$ must vanish an odd number of times during any circuit C enclosing the semifluxon. Therefore each nondegenerate eigenstate must possess at least one nodal line connecting the semifluxon to B [6]. A way to see this analytically, introducing a representation we will use later, is via the following general solution of the Helmholtz equation in (2.4) with the continuation rule (2.6), using polar coordinates $\{\rho, \mu\}$ (figure 1) centred on the semifluxon:

$$f(\rho,\mu) = \sum_{n=0}^{\infty} J_{n+\frac{1}{2}}(k\rho)\left(c_n \cos\left\{(n+\tfrac{1}{2})\mu\right\} + s_n \sin\left\{(n+\tfrac{1}{2})\mu\right\}\right)$$
$$= \sum_{n=0}^{\infty} J_{n+\frac{1}{2}}(k\rho) e_n \cos\left\{(n+\tfrac{1}{2})\mu + \chi_n\right\}. \tag{2.7}$$

For small $\rho$, that is, close to the semifluxon, the Bessel function with $n=0$ dominates, and

$$f \approx (k\rho)^{1/2} e_0 \cos\left(\tfrac{1}{2}\mu + \chi_0\right). \tag{2.8}$$

This vanishes when $\mu = \mu_{nodal} = \pi - 2\chi_0$, which is therefore the direction in which the nodal line emerges from the semifluxon. By continuity, the line cannot come to an end, so it must reach B. Figure 2 shows the phase contours for a singlevalued $\psi$, and the nodal lines of $\psi$ and $f$. As shown in Appendix A, each eigenstate pulls the semifluxon with force directed along the nodal line.

### 3. Degeneracies are inevitable for semifluxons [8]

Equation (2.6) is a real continuation rule, and since the equation and boundary condition (2.4) are also real too, the operator determining $f$ is real (false T symmetry). Therefore we can expect degeneracies as the flux position $R$ varies.

Now comes the central point. For a semifluxon we can say more: degeneracies are not only permitted but compulsory [8]. To show this, we start by imagining the semifluxon outside B. Then circuits C of the position variable $r$ inside B cannot enclose the flux, so $f$ is a singlevalued function of $r$. But neither $\psi$ nor $f$ is singlevalued under continuation of $R$ round a circuit $C_R$ enclosing B; these functions change sign, because by transforming to a reference frame in which the semifluxon is fixed this appears as the AB phase acquired by transporting B around it [10]. Now shrink $C_R$ so that it lies just inside B. By continuity, the sign change of $\psi$ and $f$ persists. Thus we have the situation where continuation of the parameters $R$ specifying the real symmetric operator O induces a sign change. This can be regarded as a geometric phase, implying [10] the existence, for each eigenstate, of at least one degeneracy inside $C_R$ – in general, an odd number of them. Alternatively stated, there must be at least one position of the semifluxon inside B for which the state is degenerate.

## 4. Circle AB billiard degeneracies

When B is a circle, we can without loss of generality choose unit radius and the semifluxon on the positive $X$ axis. Then the eigenstates (2.7) must be even or odd in $\mu$, and labelled by $c_n$ or $s_n$ respectively. To apply the boundary condition at $r=1$, we transform (2.7) from polar coordinates $\{\rho,\mu\}$, centred on the flux (figure 1), to $\{r,\theta\}$, centred on $r=0$, using the addition formula for Bessel functions [11], For the even states, we need

$$\cos\{(n+\tfrac{1}{2})\mu\}J_{n+\tfrac{1}{2}}(kr) = \sum_{m=-\infty}^{\infty} \cos\{(m+n+\tfrac{1}{2})\phi\}J_{n+m+\tfrac{1}{2}}(kr)J_m(kR). \quad (4.1)$$

Thus the even states (2.7) are

$$\begin{aligned}
f_e(r,\theta) &= \sum_{n=0}^{\infty} c_n \sum_{s=-\infty}^{\infty} \cos\{(s+\tfrac{1}{2})\theta\} J_{s+\tfrac{1}{2}}(kr) J_{s-n}(kR) \\
&= \sum_{n=0}^{\infty} c_n \sum_{s=0}^{\infty} \cos\{(s+\tfrac{1}{2})\theta\} \Big[ J_{s+\tfrac{1}{2}}(kr) J_{s-n}(kR) \\
&\quad - (-1)^{s+n} J_{-s-\tfrac{1}{2}}(kr) J_{s+n+1}(kR) \Big].
\end{aligned} \quad (4.2)$$

For all $\theta$, $f$ must vanish when $r=1$, giving a set of homogeneous linear equations for the coefficients $c_n$. Their solution requires the vanishing of a determinant, which is the quantum condition for the energy levels $k_n^2(R)$. Including the analogous condition for the odd solutions, we obtain

$$\det{}_{s,n} M_{\substack{\text{even}\\\text{odd}}}(k;R) = 0 \quad (0 \leq s,n < \infty), \quad (4.3)$$

where

$$M_{\substack{\text{even}\\\text{odd}}}(k;R) = J_{s+\frac{1}{2}}(kr) J_{s-n}(kR) \mp (-1)^{s+n} J_{-s-\frac{1}{2}}(kr) J_{s+n+1}(kR). \quad (4.4)$$

These equations are easy to solve numerically, giving the spectrum shown in figure 3. It is clear that at $R=0$ (semifluxon at the centre of the circle) each odd level is degenerate with an even one. These levels are the zeros $k_{n,s}$ of the half-integer Bessel functions $J_{n+1/2}(k)$. It seems as though the third and fourth odd levels cross near $R=0.2$, but in fact this is an avoided crossing, as the inset shows. As $R \to 1$, the levels approach those for the circle without flux, that is, the zeros of the integer Bessel functions $J_n(k)$, which are degenerate for $n \neq 0$ and nondegenerate for $n=0$. (The apparent inconsistency that the ground state for $R=1$ appears odd, rather than even, is resolved by careful reinstatement of the magnetic phase factor in (2.3).)

## 5. Perturbed circle AB billiard degeneracies

To compute energy levels for semifluxon billiards with a general boundary, we first represent B parametrically, in terms of an angle $0 \leq \phi \leq 2\pi$, as $\{x(\phi), y(\phi)\}$. Then the radial and angular variable occurring in the flux-centred representation (2.7) are given (cf figure 1 with *r* on B) by

$$\rho(\phi) \exp(i\mu(\phi)) = x(\phi) + iy(\phi) - X - iY. \quad (4.1)$$

Now choose angles

$$\phi_m = \frac{m\pi}{N} \quad (1 \leq m \leq 2N), \tag{4.2}$$

generating 2N points on B. The wave $f(\rho(\phi_m), \mu(\phi_m))$ given by (2.7) must vanish for all points $m$, and truncating the sum in (2.7) leads to a set of linear equations for the coefficients $c_n, s_n$. The condition for their solution is

$$\det M_{m,n}(k; X, Y) = 0 \quad (1 \leq m, n \leq 2N), \tag{4.3}$$

where

$$M_{m,n}(k; X, Y) = \left.\begin{array}{l} J_{n-\frac{1}{2}}(k\rho(\phi_m; X, Y)) \cos\{(n-\tfrac{1}{2})\phi_m\} \\ (1 \leq n \leq N, 1 \leq m \leq 2N) \end{array}\right\} \\ = \left.\begin{array}{l} J_{n-N-\frac{1}{2}}(k\rho(\phi_m; X, Y)) \sin\{(n-N-\tfrac{1}{2})\phi_m\} \\ (N+1 \leq n \leq 2N, 1 \leq m \leq 2N) \end{array}\right\}. \tag{4.4}$$

In the limit $N \to \infty$, (4.3) is an eigenconditions determining the energy levels $k_n^2$. In the computations of low-lying levels to follow, sufficient accuracy is achieved with $N \leq 20$; in most cases $N=10$ suffices.

Now we must specify the boundary. To avoid confusion with degeneracies between crossings of eigenstates belonging to different geometric symmetry classes (e.g. the even and odd states in figure 3), we choose B with no symmetry. A simple three-parameter family of such boundaries, with each B labelled by $a_2, a_3, \sigma$ and given parametrically by an angle $\phi$, is the following cubic conformal transformation of the unit circle:

$$x(\phi) + iy(\phi) = \exp(i\phi) + a_2 \exp(2i\phi) + a_3 \exp(3i\phi + i\sigma) \quad (4.5)$$

(families containing only $\phi$ and $2\phi$ are too simple because they possess reflection symmetry). In this section, we choose

$$a_2 = 0.015, a_3 = 0.05, \sigma = \tfrac{1}{3}\pi, \quad (4.6)$$

A test of the computation scheme is to fix the semifluxon position $X,Y$ and compare the spectral staircase $\mathcal{N}(E)$ (number of levels with $E_n < E$) with the smoothed staircase given by the Weyl approximation including the terms corresponding to the billiard area $A$, perimeter length $L$, and the topology and flux-dependent terms [12]. For semifluxons, this is

$$\mathcal{N}_{smoothed}(E) = \frac{AE}{4\pi} - \frac{L\sqrt{E}}{4\pi} + \frac{1}{12}. \quad (4.7)$$

This is a stringent test because missing or superfluous levels would be signalled by systematic deviations between $\mathcal{N}(E)$ and $\mathcal{N}_{smoothed}(E)$. As figure 4 shows, the agreement is excellent.

To seek degeneracies, we scan the parameter space $X$, $Y$ as illustrated in figure 5a. For fixed $Y$, the zero contours of the determinant in (4.3) in the $\{X,k\}$ plane are the energy levels (figure 5b). Then, changing $Y$, concentrating on values where levels come close together (figure 5c), we home in on the degeneracies. We checked each candidate for a degeneracy by radially scanning $\sqrt{X^2+Y^2}$ for fixed azimuth $\arg(X+iY)$. In this way we located all the degeneracies involving the first three levels, as illustrated in figure 6 and listed in Table 1.

| levels $n, n+1$ | flux position $x, y$ | eigenvalue $k$ |
|---|---|---|
| 1,2 | 0.16, –0.03 | 3.05 |
| 2,3 | 0.71, –0.45 | 3.82 |
| 2,3 | –0.40, 0.44 | 3.82 |
| 3,4 | –0.08, 0.08 | 4.34 |
| 3,4 | 0.38, 0.10 | 4.40 |
| 3,4 | 0.18, –0.3 | 4.38 |

Table 1. Degeneracies involving the first three levels

In the parameter space of semifluxon positions, the lowest level has one degeneracy (with level 2 of course), level 2 has three degeneracies (one with level 1 and two with level 3), and level 3 has five degeneracies (two with level 2 and three with level 4). This pattern is consistent with the theoretical predictions of section 3, As will be shown in the next section, other patterns can also occur.

## 6. Dance of the degeneracies

Degeneracies can themselves degenerate. Here we explore two ways in which this can happen. The first way is involves degeneracies of more than two levels. A coincidence of three levels has codimension 5 (the number of parameters in a 3x3 traceless real symmetric matrix). This implies that for a single semifluxon and fixed B there will in general be no position $X, Y$ for which an eigenstate is triply degenerate: in addition to $X$ and $Y$, three boundary shape parameters must be varied. Another possibility is to fix B and thread the billiard with more semifluxons. The minimum number necessary to find a degeneracy is three. Then there would be one parameter too many: in the six-parameter space $X_1, Y_1, X_2, Y_2, X_3, Y_3$, we expect the degeneracy locus to be a line. Generalizing, coincidence of $N$ levels has codimension $(N+2)(N-1)/2$, and could be achieved for fixed B by exploring the position space of $\text{Int}\{N(N+1)/4\}$ semifluxons.

The second way involves the same two levels, and a pair of degeneracies corresponding to different semifluxon positions $\boldsymbol{R}$. These can be made to collide and annnihilate, or be born spontaneously, by varying a single boundary shape parameter. This process can be illustrated by the 2x2 matrix

$$M = \begin{pmatrix} f(X,Y,Z) & g(X,Y,Z) \\ g(X,Y,Z) & -f(X,Y,Z) \end{pmatrix}, \qquad (6.1)$$

in which we can regard $X$ and $Y$ as the position of the semifluxon as before, and $Z$ as parameterising the shape of B. Eigenvalues correspond to $X, Y, Z$ for which $f$ and $g$ are simultaneously zero, or, equivalently, as zeros of the complex function $f+ig$, with degeneracies corresponding to coinciding zeros. Explicitly,

$$M = \begin{pmatrix} Z - X^2 & Y \\ Y & -Z + X^2 \end{pmatrix} \qquad (6.2)$$

has two degeneracies for $Z>0$, at $X = \pm\sqrt{Z}, Y = 0$, coinciding when $Z=0$, and none for $Z<0$.

We have located such an event in our cubic conformal billiard family with B given by (4.5), fixing $a_3$ and $\sigma$ as in (4.6) and varying $a_2$, and studying degeneracies between levels 3 and 4. We know that for $a_2=0.015$ there are three degeneracies (cf figures 6). For $a_2=0$ there cannot be three, because this is incompatible with the symmetry of B in this case, which is that of an ellipse: reflection symmetry about the perpendicular diretions $-\pi/6$ and $+\pi/3$. And computation shows that there there are five degeneracies (figure 7a), symmetrically arranged with the correct symmetry. Therefore there must be a collision event in the interval $0<a_2<0.015$. To find it, we chose increasing values of $a_2$, and for each of them we plotted the zero contours of the determinant in the $X$, $Y$ plane for a series of values of $k$, seeking the degeneracies, where the loop sections of the double-cone energy level surfaces shrink to a point; this is a laborious procedure, because the cones expand and merge over very small ranges of $k$. Figure 7 shows the result: the collision is the annihilation of two degeneracies, that numerics indicate happens close to $a_2=0.007108$.

For this collision event, all parameters $a_2$, $a_3$, $X$, $Y$ are small, suggesting the possibility of an analytical description, based on perturbation theory from the circular billiard with the semifluxon at the centre. We have not pursued this.

**Acknowledgments**

MVB's research is supported by the Leverhulme Trust.

**Appendix A. Force on a semifluxon**

There must be a force on the semifluxon, at position $\boldsymbol{R}$, because the energy $E(\boldsymbol{R})$ depends on $\boldsymbol{R}$. For the eigenstate $|\psi(\boldsymbol{R})\rangle$, the force is, after applying the Hellman-Feynman theorem,

$$F(R) = -\nabla_R E(R) = -\langle \psi(R) | \nabla_R H(R) | \psi(R) \rangle, \qquad (A.1)$$

involving the force operator $-\nabla_R H(R)$. Keating and Robbins [13] show that in order to express the expectation value (A.1) in terms of the behaviour of the wavefunction at the flux it is necessary to take careful account of the singular nature of the force operator. We now show that when applied to the semifluxon bound states we are considering here, their formulas imply that the force is directed along the nodal line described in section 2.

In flux-centred coordinates (figure 1), they write the wavefunction (formula (17) of [13]) as

$$\psi(\rho,\mu) = \sum_{m=-\infty}^{\infty} \psi_m(\rho) \exp(in\mu) \qquad (A.2)$$

(cf. (2.7)), and then, for general flux $\alpha$, the force, expressed in a complex notation, is (formula (18) of [13])

$$\langle \psi | (F_x + iF_y) | \psi \rangle = $$
$$-2\pi C \sum_{m=-\infty}^{\infty} \left[ \psi_{m+1}^*(\rho)(m+1-\alpha) \left( \psi_m'(\rho) - \frac{(m-\alpha)}{\rho} \psi_m(\rho) \right) \right]_{\rho=0}. \qquad (A.3)$$

Now $\psi_m(\rho) \propto \rho^{|m-\alpha|}$ [14], so for the semifluxon case $\alpha=1/2$, the dominant behaviour at the origin comes from $m=0$ and $m=1$, and without loss of generality we can write, consistent with (2.8),

$$\psi_0(\rho) = C\sqrt{\rho}, \quad \psi_1(\rho) = C\sqrt{\rho}\exp(2i\chi_0), \tag{A.4}$$

where $C$ is a constant, thus regenerating the wavefunction (2.8) with the nodal line issuing from the semifluxon in the direction $\mu_{nodal} = \pi - 2\chi_0$. In (A.3) only the term $m=0$ survives at $\rho=0$, leaving

$$\langle\psi|(F_x + iF_y)|\psi\rangle = -\pi C\exp(-2i\chi_0) = \pi C\exp(i\mu_{nodal}). \tag{A.5}$$

Therefore the force acts to shorten the nodal line, which resembles an elastic string.

If there are several nodal lines, then, in addition to the force (A.5), acting on each of them from the eigenstate, they exert forces on each other, as discussed previously [15].

**Figure captions**

**Figure 1**. Geometry and coordinate for AB billiard wavefunction and semifluxon.

**Figure 2**. Nodal lines (thick), and phase contours (thin, with phase values indicated), for the 5th energy eigenstate $\psi$ of the AB semifluxon 'africa' billiard with boundary (4.5) with $a_2=a_3=0.2$, $\sigma=\pi/3$ (after figure 2d of [6]

and with the same choice of gauge).

**Figure 3**. Energy levels of the circle billiard as a function of the distance $R$ if the semifluxon from the centre; full curves: even levels, dashed curves: odd levels. Inset: magnification showing avoided crossing of odd levels.

**Figure 4**. Test of level computation scheme for the billiard (4.5, 4.6) with semifluxon at $X=Y=0$. Full curve: spectral staircase, with levels computed from (4.3) with $2N=40$ boundary points; dashed curve: smoothed staircase (4.7) (for the billiard (4.6), the area is $A=3.34328$ and the perimeter is $L=7.10123$).

**Figure 5**. Sample scan lines (a) of semixfluxon position $X$ for fixed $Y$, in the billiard (4.5, 4.6), to determine (b) the levels $k$ as a function of $X$. There is a near-degeneracy of levels 2 and 3 for $Y=+0.4$ (Table 1), and two degeneracies of levels 3 and 4 near $Y=0$, located by zooming in (c).

**Figure 6**. Semifluxon locations for which levels $\{n,n+1\}$ are degenerate for the billiard (4.5, 4.6) for $n=1,2,3$.

**Figure 7**. Collision of two degeneracies between levels 3 and 4 for the billiard (4.5) with $a_3=0.08$, $\sigma=\pi/3$ and (a,b) $a_2=0$, (c) $a_3=0.005$. (d) $a_3=0.007$, (e,f) $a_3=0.0072$. The collision occurs between (d) and (e). (b) and (e) are magnifications of (a) and (f).

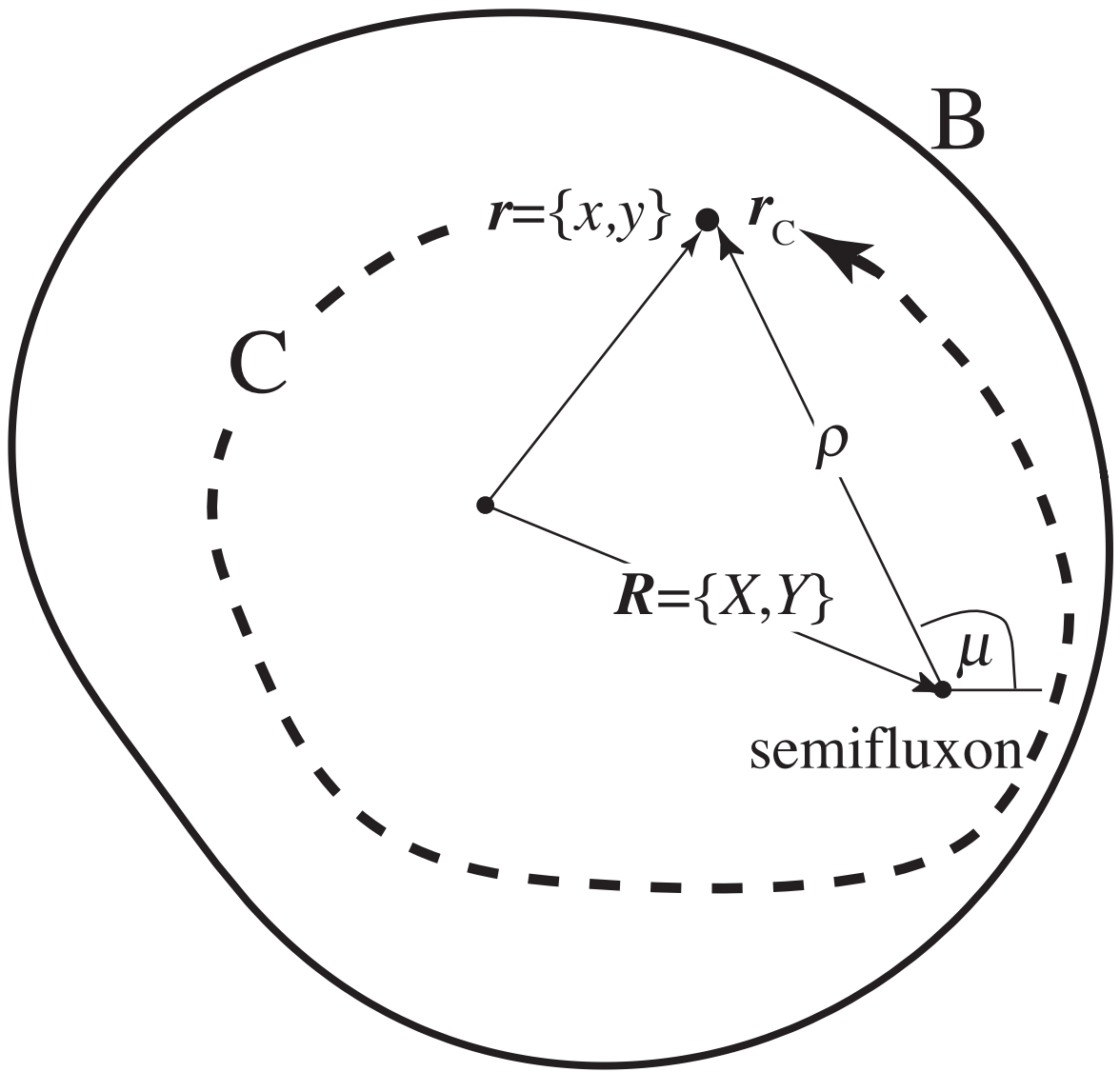

figure 1

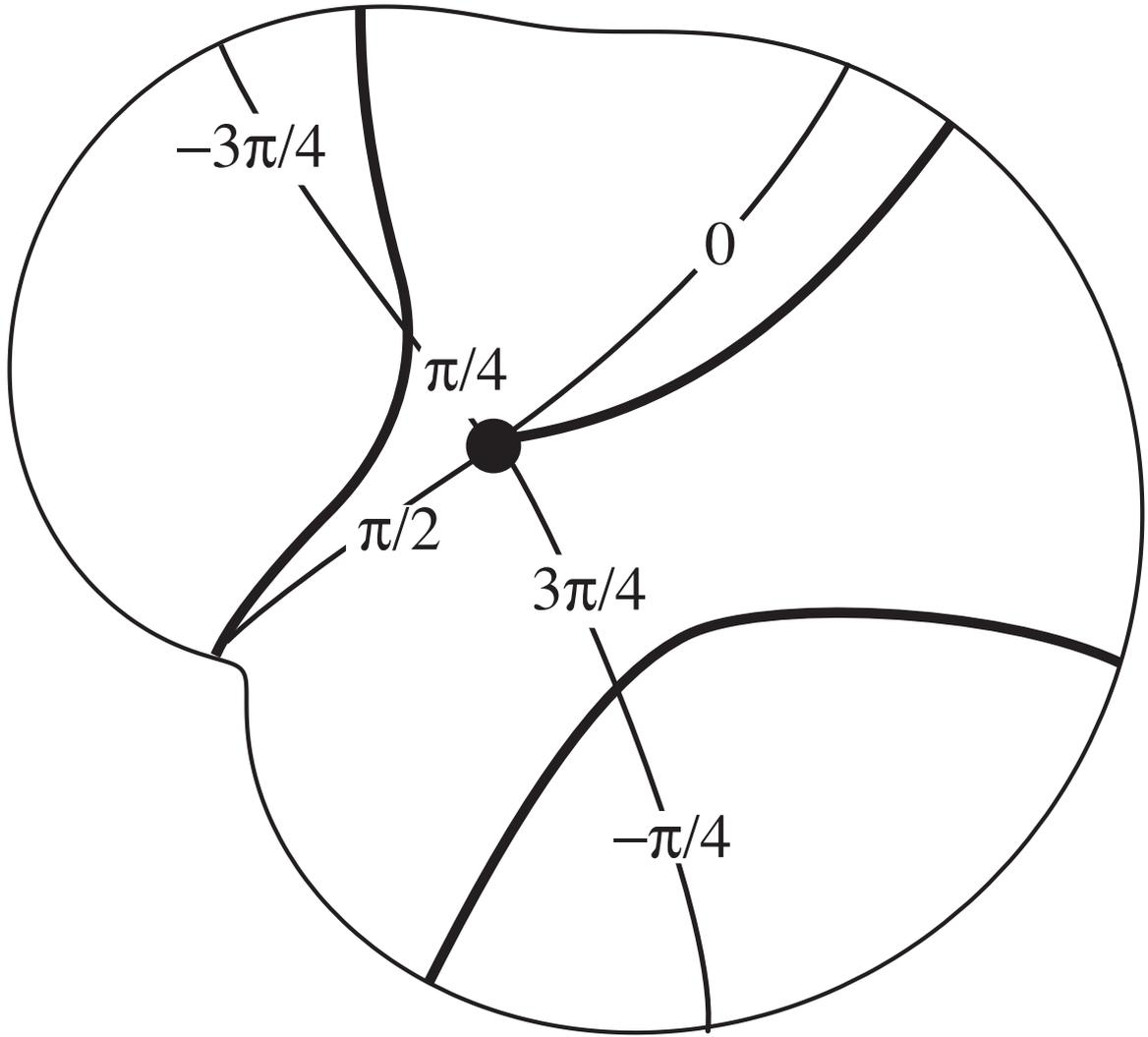

figure 2

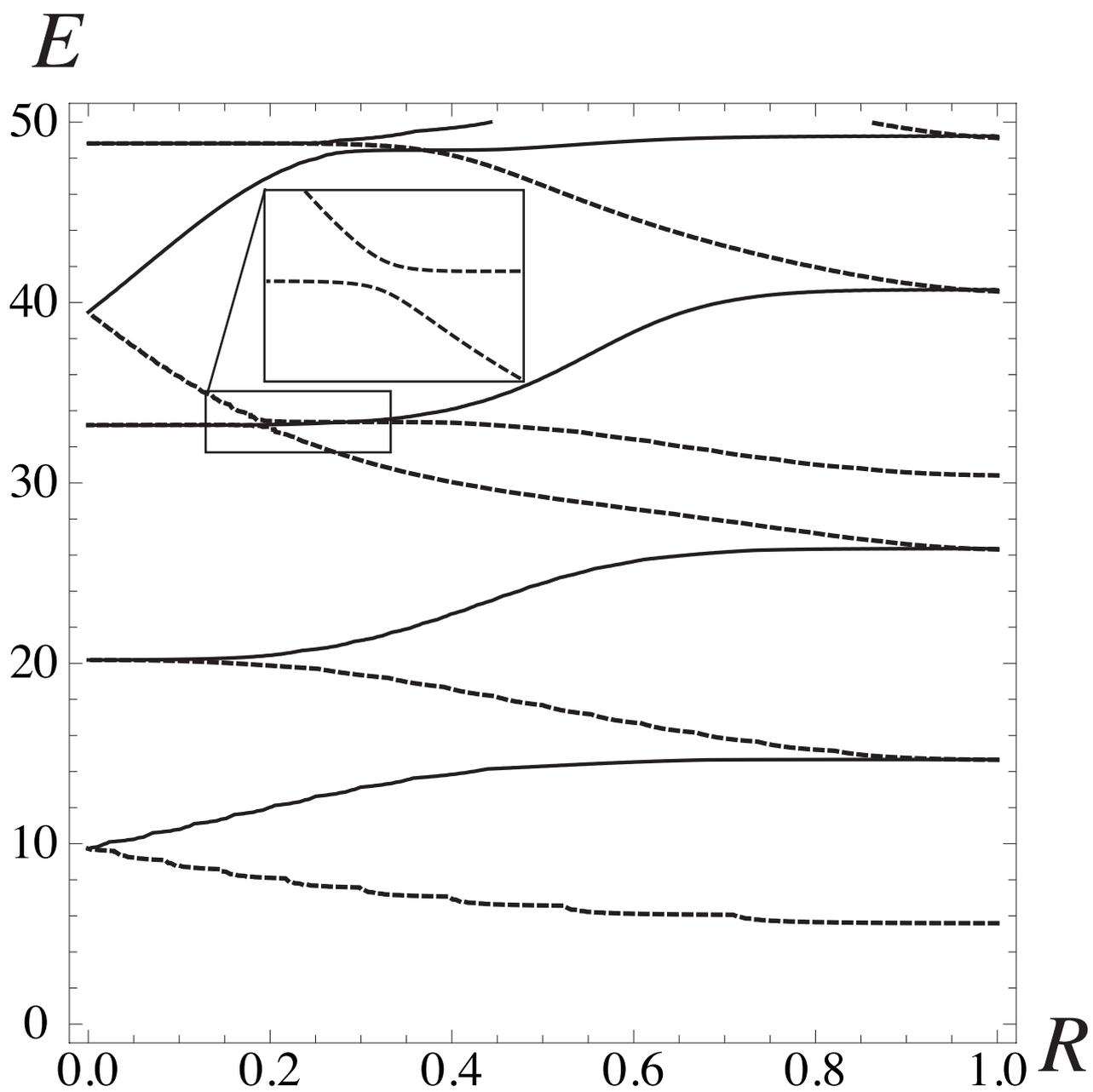

figure 3

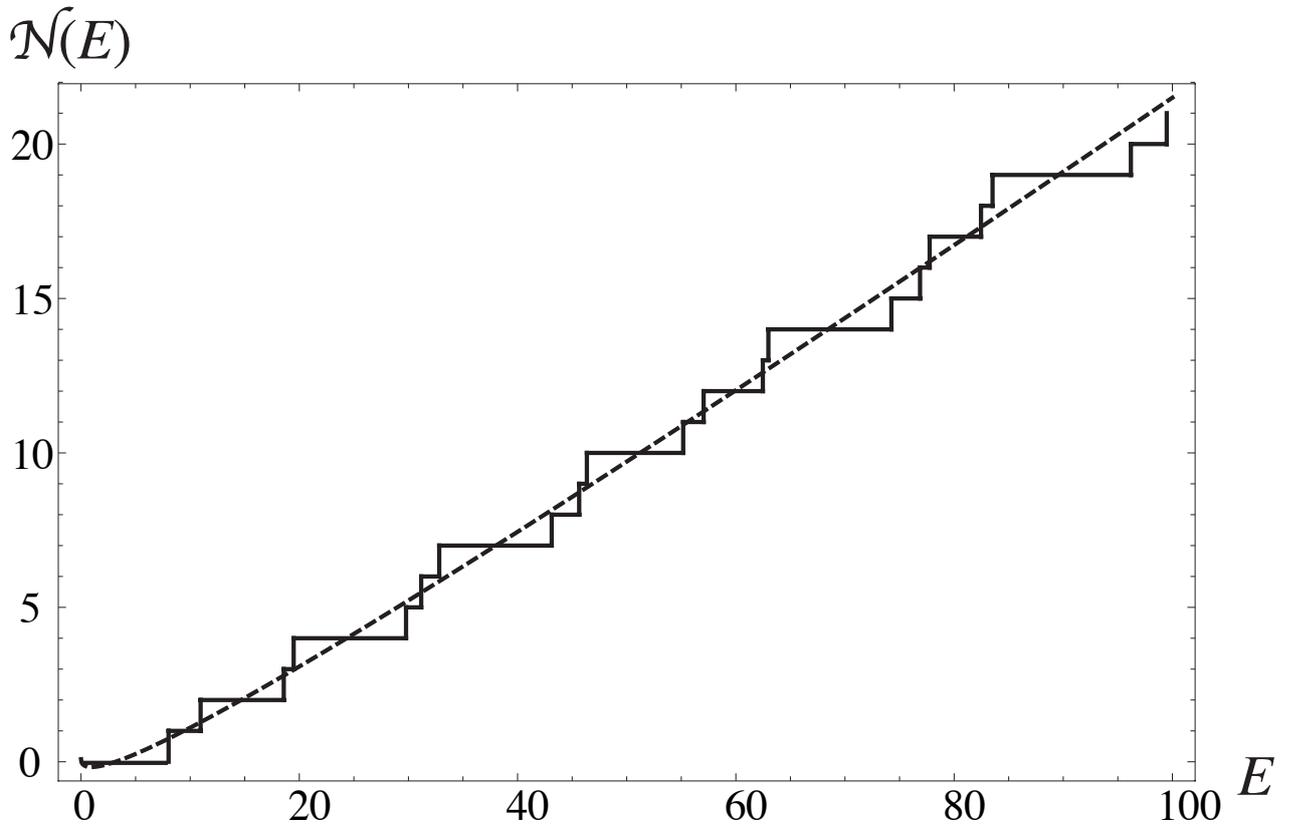

figure 4

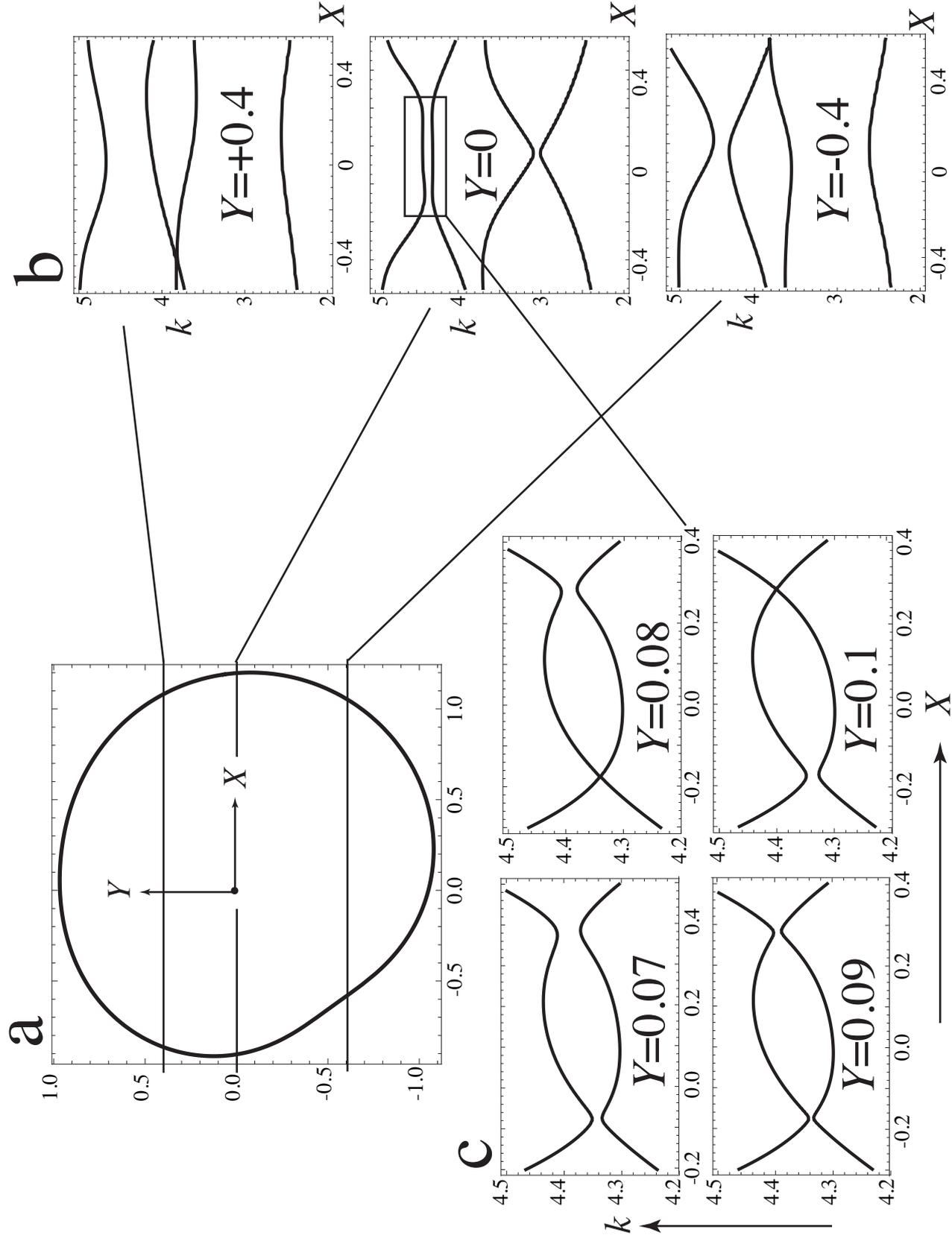

figure 5

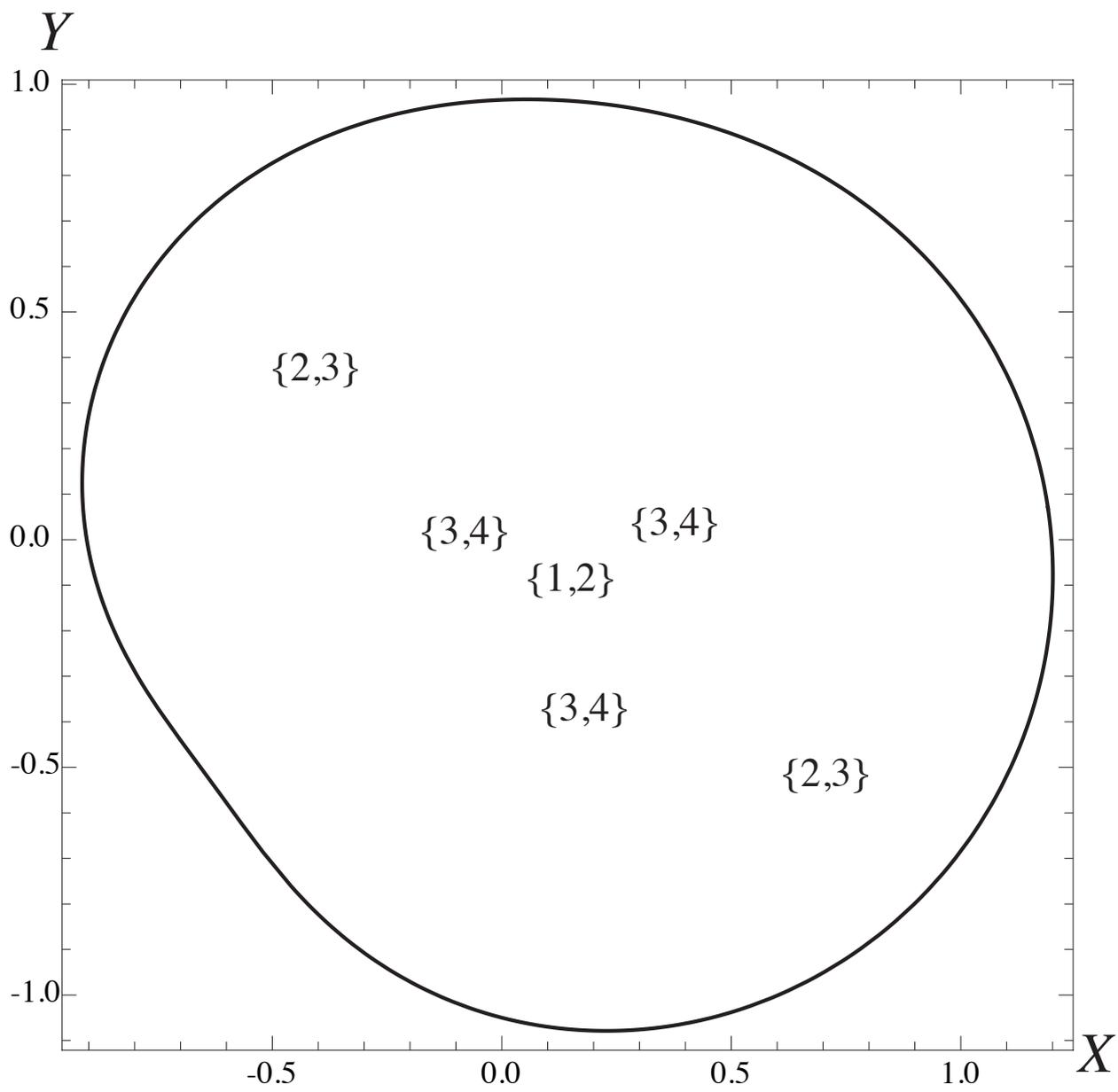

figure 6

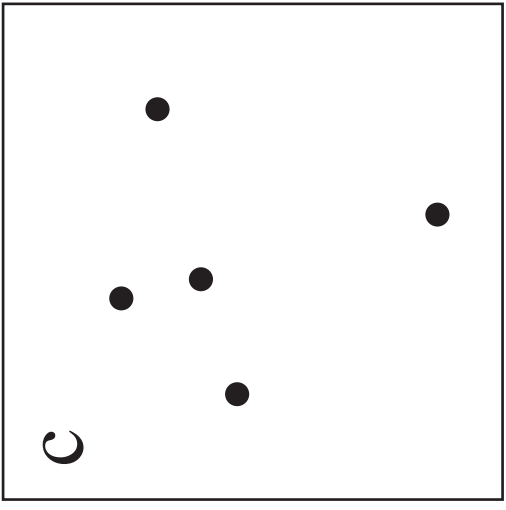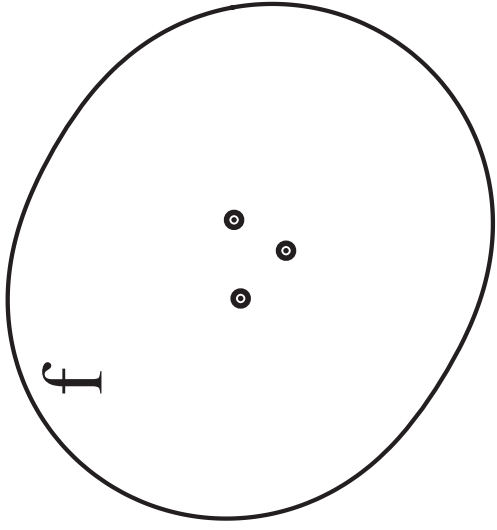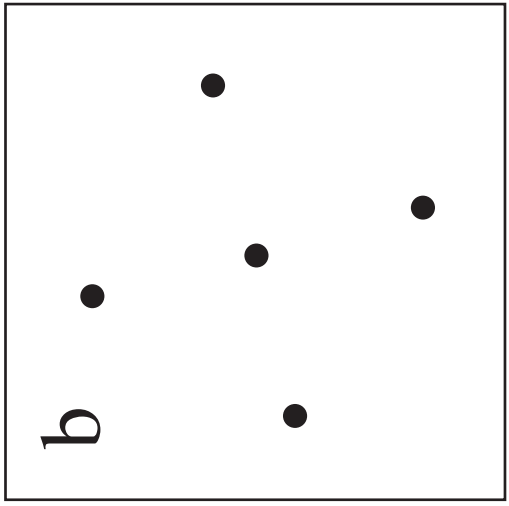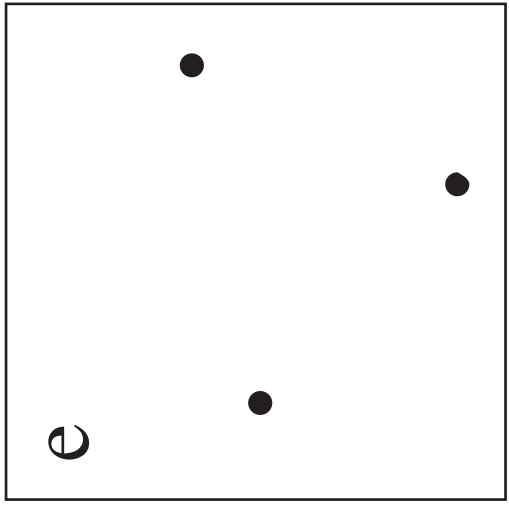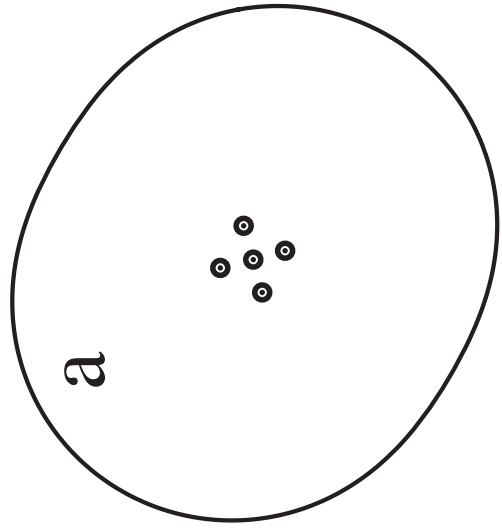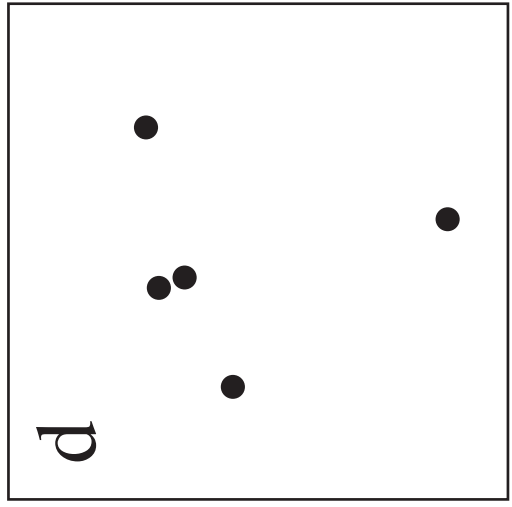

figure 7